\documentclass{article}
\usepackage{hip-artc}
\usepackage{graphicx}
\usepackage[center]{subfigure}
\usepackage{amsmath}
\usepackage{amssymb}
\volnumber{} \issuenumber{} \edyear{2004}                                
\frompage{000} \topage{000}                                              
\recrevdate{...}
\def\rightmark{Gluon Emission of Heavy Quarks: Dead Cone Effect}
\def\leftmark{R. Thomas, B. K\"{a}mpfer, G. Soff}

\title{Gluon Emission of Heavy Quarks: Dead Cone Effect}
\authors{
{R. Thomas$^{1}$, B. K\"{a}mpfer$^1$, G. Soff$\mspace{3mu}^{2}$%
\index{One, A.}
\index{Two, A.}
}\\[2.812mm]
{\normalsize
\hspace*{-8pt}$^1$ Forschungszentrum Rossendorf, PF 510119,
01314 Dresden, Germany\\[0.2ex]
\hspace*{-8pt}$^2$ Institut f\"{u}r Theoretische Physik, TU Dresden,
01062 Dresden, Germany
}}

\abstract{The lowest-order induced soft gluon radiation processes
of heavy quarks have been analyzed to quantify the dead
cone effect. This effect is most likely expected to suppress
significantly the energy loss of charm quarks passing an amorphous
color-neutral deconfined medium, as has been concluded from recent
experiments at RHIC.}


\makeindex
\begin{document}

\maketitle \setcounter{page}{1}
\section{Introduction}\label{intro}

Radiative energy loss of quarks traversing strongly interacting
matter is considered as a promising probe of the quark gluon
plasma. Indeed, the recent heavy-ion collision experiments at RHIC
show a distinct suppression of transverse momentum spectra of
hadrons composed of light quarks \cite{RHIC} thus evidencing the
transient creation of a medium with large stopping power caused by
a high gluon density. In contrast, open charm mesons
\cite{Averbeck} seem to suffer only a tiny, if any, energy loss
\cite{Gallmeister}. As noted in \cite{Dok01} prior to this
experimental finding, the gluon emission pattern of heavy quarks
experiences a modification, known from jet physics as dead cone
effect, which reduces the radiative energy loss. While large
efforts are devoted to the study of the energy loss of light
quarks at asymptotic energies (cf.\
\cite{BDMPS,Gyu00a,GLV,Wiedemann,Arnold,B.Mueller,X.N.Wang,reviews,Gyu94} 
and further references therein), the treatment of heavy quarks 
\cite{U.A.W.,Gyulassy_NPA} is in progress.

Similar to the fairly complete treatment of the radiative energy
loss of light high-energy quarks we consider here the lowest-order
induced gluon emission of a heavy quark. It is our goal to take
exactly into account the finite mass, the finite energy and
correct kinematics (i.e. not only soft gluon emission), the full
color algebra and all spin/polarisation effects. In doing so we
restrict ourselves to the single scattering and one-gluon emission
process. The importance of the single scattering process is
emphasized by various aspects. Among of them data from RHIC, which
suggest thin plasmas, imply that only a few scatterings are taking
place; single scattering is the first, and according to \cite{GLV}
the most important. The existence of an analogue to the Landau-Pomeranchuk-Migdal 
(LPM) effect in QCD, which reduces the effective number of scatterings further,
is another indication of the significance of this process.
Moreover the fast convergence of the opacity expansion \cite{GLV},
where the first order describes the radiative energy loss
considerably well, suggests the relevance of this basic energy
loss process. The discussion of the QCD analog of the Ter-Mikaelian (TM) effect
\cite{TM} is also based on the lowest order in opacity 
\cite{Magdalena}\footnote{The results of \cite{Magdalena} confirm
the conjecture \cite{TM} that the asymptotic gluon mass, which is also
important for describing phenomenologically the quark-gluon plasma
\cite{Peshier}, determines the TM effect.}.

The article is organized as follows. In Section 2 we describe the
potential model and the concept of radiation amplitude and dead
cone factor. Numerical results are then discussed in Section 3.
The validity of the potential model is commented on in Section 4.
The last part summarizes conclusions.

\section{Single Scattering in the Potential Model}\label{techno}

\begin{figure}
\subfigure[$M_{1,0,0}$ \newline (pre emission)]{
\includegraphics[width=2cm, angle=90]{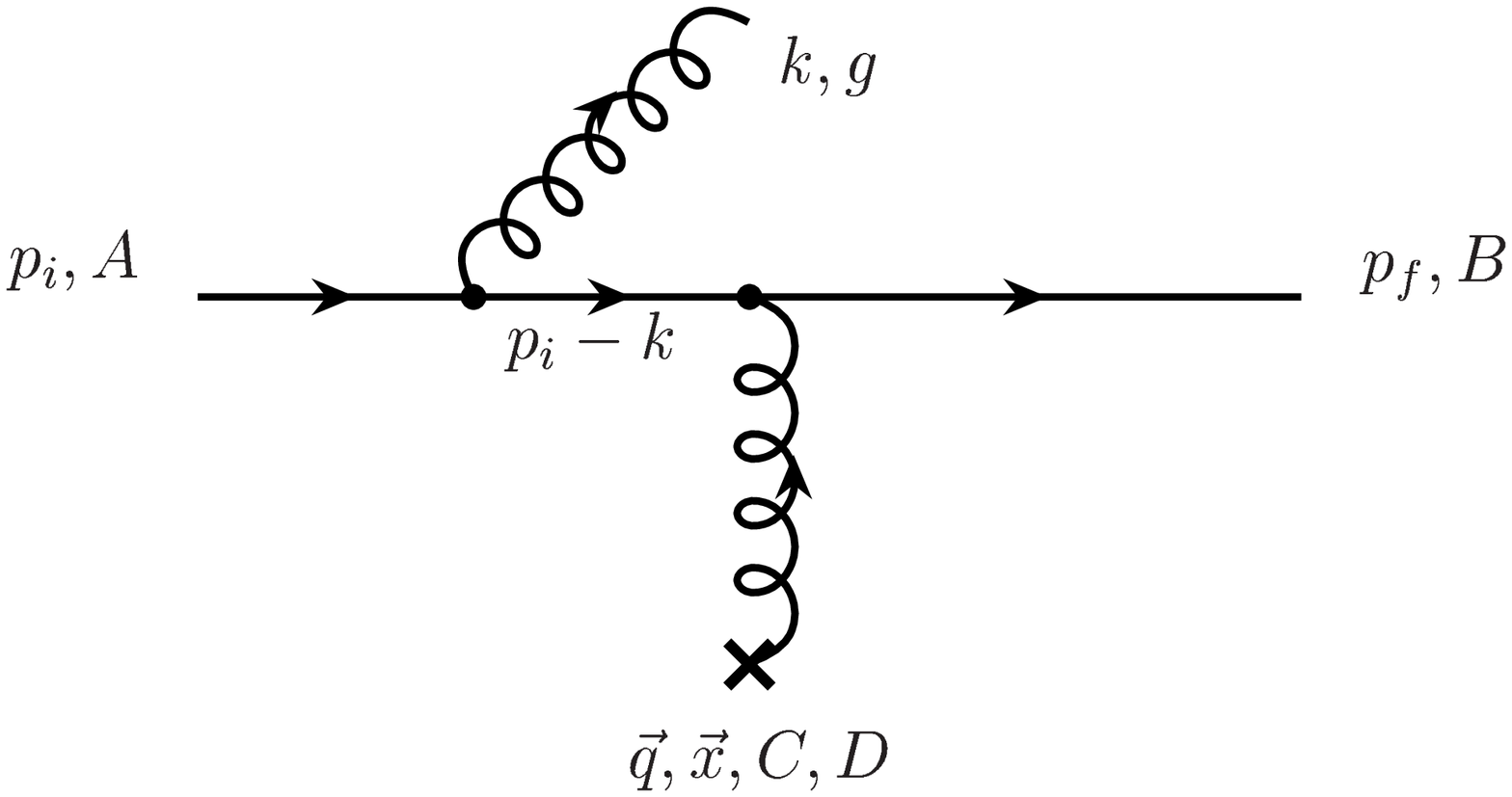}}
\subfigure[$M_{1,1,0}$ \newline (post emission)]{
\includegraphics[width=2cm, angle=90]{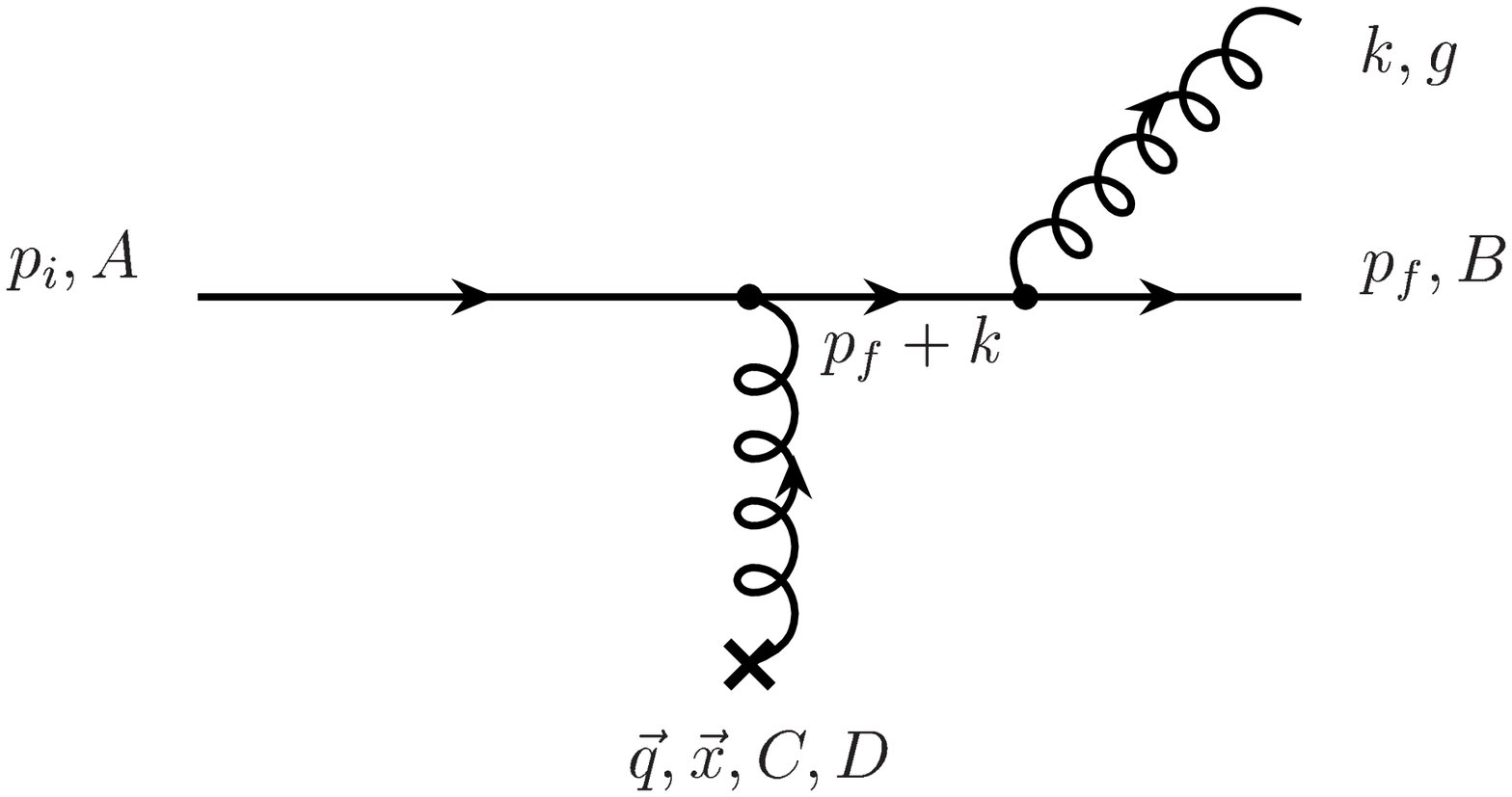}}
\subfigure[$M_{1,0,1}$ \newline (three gluon)]{
\includegraphics[width=2cm, angle=90]{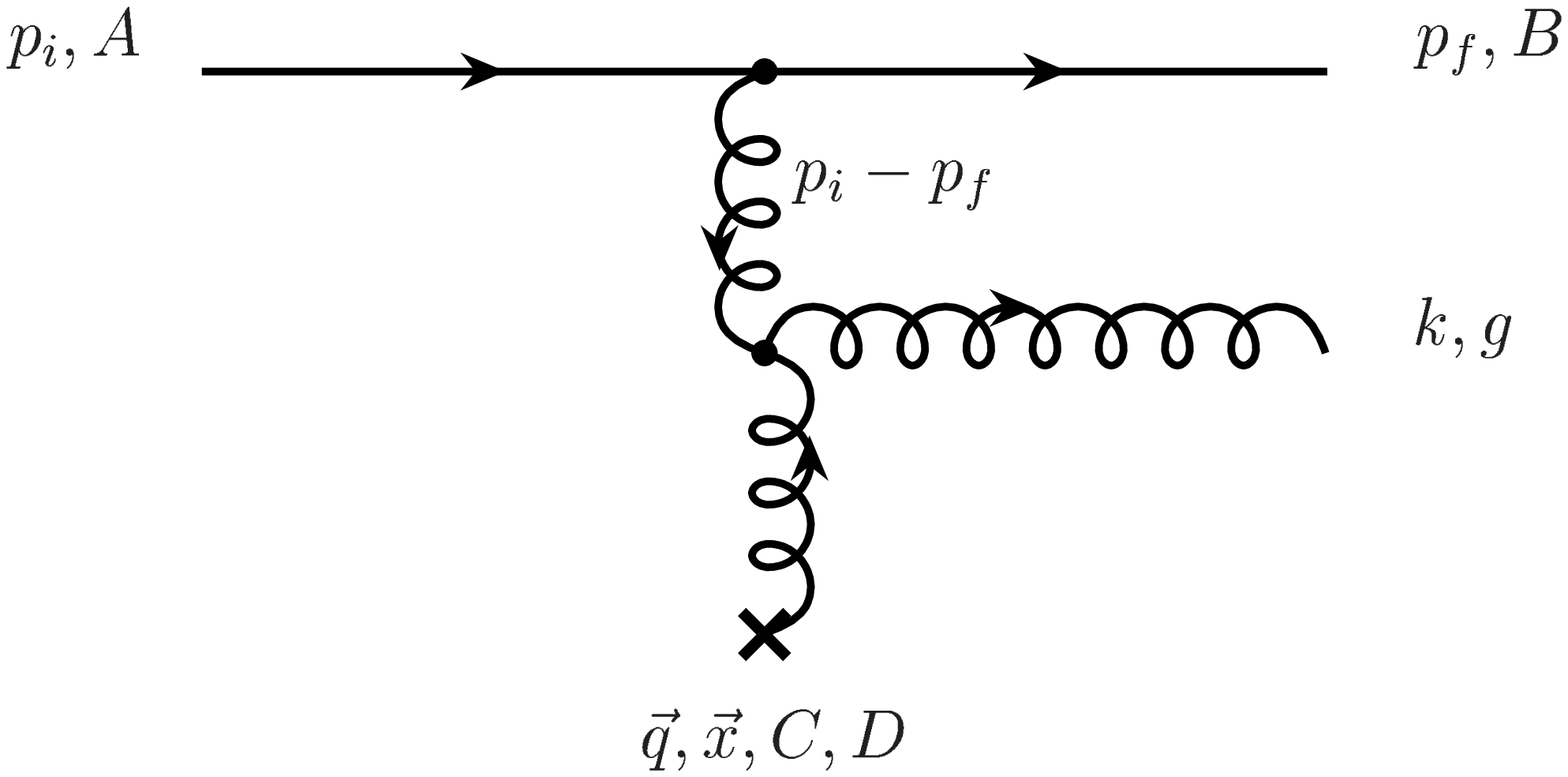}}
\vspace*{-0.5cm} \caption[]{Diagrams of single scattering with
one-gluon emission classified by the notation scheme $M_{n,m,l}$
of \cite{Gyu00a}.} \label{diagrams}
\end{figure}

Following \cite{GB} we consider the processes depicted in Fig.~1.
The crosses stand for external potentials which, according to
\cite{Gyu94}, represent the medium.  
A color charge, here a quark, moving through a background
field of quarks and gluons is mimicked by  scattering on target
particles modeled by static screened potentials localized at the
space points $\vec{x}_i$,
$ 
V^a_{AB}(\vec{q}\,) = g T^a_{AB}
\dfrac{e^{-i\vec{q}\vec{x}_i}}{\vec{q}\,^2 + \mu^2}, 
$ 
$\vec{q}$ is the momentum transfer and $\mu$ is the Debye
screening mass; the coupling $g$ is not varying here. The term
$T^a_{AB} = \chi^\dagger_B C^{a_1 \cdots a_m}_{b_1 \cdots b_n}
\chi_A$ stands for the original color structure of the target,
where $T^a$ are generators of $SU(3)_{color}$, $\chi$ the quark
color states and $A,B = 1 \ldots 3$ are color indices. In order to allow
for the neglect of radiation contributions from targets in the realistic situation where
the potential in Fig.~1 is substituted by a dynamical target, and thus for a realistic
model, one requires
a particular gauge choice, the $A^+$-gauge. The  gauge choice is
given by the condition $A^+ \equiv A^0 +A^z = 0$ in light-cone
coordinates.

In the soft radiation limit the conditional probability of one-gluon emission
under the condition that an elastic scattering has occurred is given by the ratio
of the inelastic and elastic differential cross sections
\begin{equation}
\mathrm{d}P = \dfrac{\overline{\mathrm{d}\sigma_{\rm inel}}}{\overline{\mathrm{d}\sigma_{\rm el}}}
= \dfrac{\overline{|M_{\rm inel}|^2}}{\overline{|M_{\rm el}|^2}} 
\, \dfrac{\mathrm{d}^3k}{2\omega(2\pi)^3},
\end{equation}
where the overline indicates an average over initial polarization
and color configurations and a sum over all final polarizations
and colors. Thus the sufficient quantity to compare different
radiation patterns in such scattering processes is the radiation
amplitude
$ 
\overline{|R|^2} = \overline{|M_{\rm inel}|^2} / \overline{|M_{\rm
el}|^2}. 
$ 
If the momentum of the gluon is small, compared to the momenta of
the other particles, a factorization into a part describing the
emitted radiation and an underlying part for the elastic
scattering can be carried out on the level of the matrix elements.

The probability, i.e.\ the radiation amplitude, for gluon emission off heavy quarks
in projectile direction is strongly reduced at radiation angles
$\theta < \theta_0$ by a dead cone suppression factor according 
to $F^2$ \cite{Dok01} with
\begin{equation}
\label{deadconefactor}
F = \dfrac{k^2_\perp}{k^2_\perp + \omega^2 \theta^2_0} = 
\dfrac{\sin^2 \theta}{\sin^2 \theta + \theta^2_0},
\end{equation}
which corrects the matrix elements derived in the approach of
Gunion and Bertsch \cite{GB} for massless, high energetic quarks.
Below $\theta_0 \equiv m / E$, where $E$ means the energy of the incident
heavy projectile quark, radiation becomes suppressed due to the
mass $m$ of the projectile, which introduces a new scale relevant 
for the radiation pattern.
However, the suppression
factor $F$ is restricted to small angles. Moreover, this estimate
is valid for soft gluons, that means it focuses on abelian
diagrams only, and cannot properly handle intermediate gluon energies.
These shortcomings can be overcome only by a numerical
computation, which is the main motivation for the present work.

\section{Numerical Results}\label{numericalresults}

\begin{figure}[h]
\center{\includegraphics[width=6.5cm,angle=90]{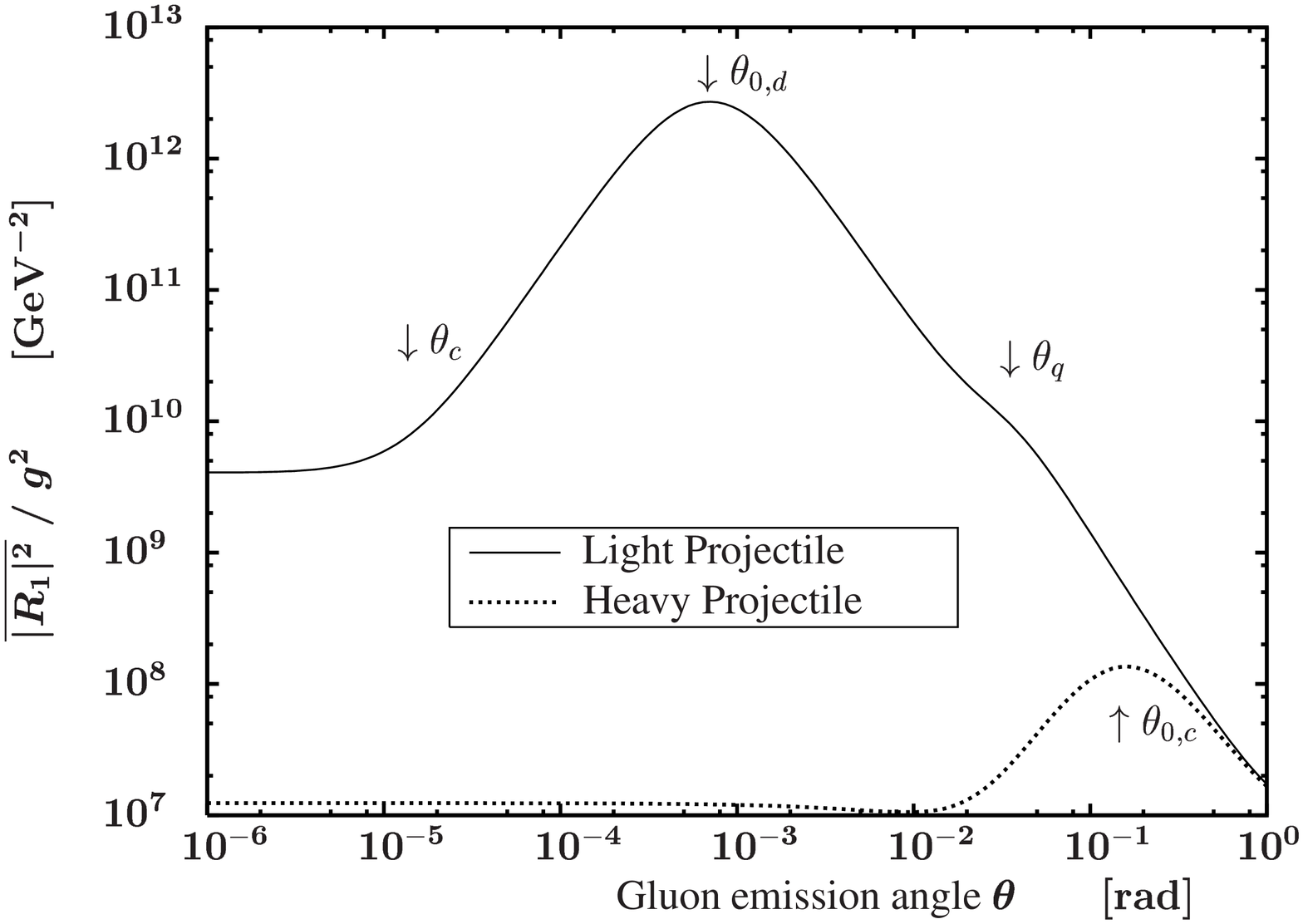}}
\vspace*{-0.5cm} \caption{Radiation amplitude $\overline{|R_1|^2}$
as a function of $\theta$ for $m < |\vec{q}_{\!\perp}|$ (light
projectile quark, $m_d = 0.007$ GeV) and
$m > |\vec{q}_{\!\perp}|$ (heavy projectile quark,
$m_c = 1.5$ GeV). Radiation is suppressed for $\theta < \theta_0$ 
due to the dead cone effect, whereby the reduction is
stronger as the quark mass grows. The chosen parameters are
$\vec{p}_i = (0,0,10)$ GeV, $\vec{p}_{f \! \perp} = (0.3,0.2)$ GeV,
$\mu = 0.5$ GeV, and $\omega = 0.001$ GeV, $\phi = \pi / 2$ for the emitted gluon.}
\label{combinedheavylightabelian}
\end{figure}

\begin{figure}[h]
\center{\includegraphics[width=6.5cm, angle=90]{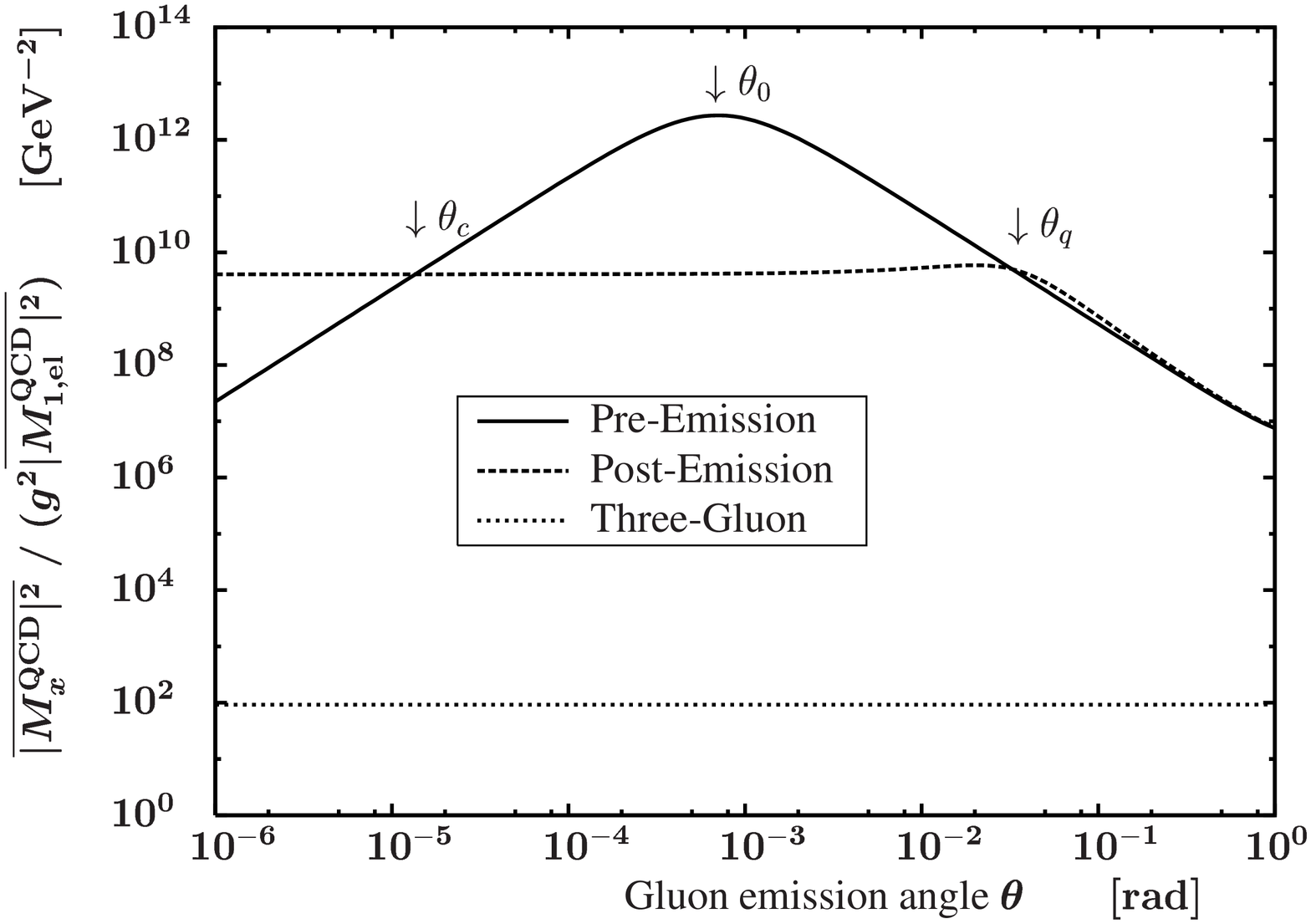}}
\vspace*{-0.5cm} \caption{Contributions of individual diagrams to
the total results in Fig.~\ref{combinedheavylightabelian} 
for the light quark scenario. The
three-gluon contribution is suppressed by orders of magnitude and
so the full QCD result does not deviate from the abelian QCD
radiation amplitude.} \label{qcdsinglepotentiallightB}
\end{figure}

\begin{figure}[h]
\center{\includegraphics[width=6.5cm, angle=90]{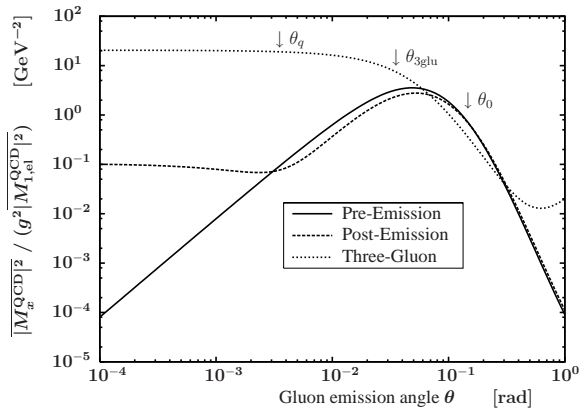}}
\vspace*{-0.5cm} \caption{Contributions of individual diagrams for a situation with a
dominating three gluon vertex diagram.
The parameters are chosen for $\theta_{\rm 3glu} < 1$, 
so that $\theta_{\rm 3glu} < \theta$ is possible,
$m_c = 1.5$ GeV, $\vec{p}_i = (0,0,10)$ GeV,
$\vec{p}_{f \! \perp} = (0.03,0.02)$ GeV,
$\mu = 0.05$ GeV, $\omega = 1$ GeV, $\phi = \pi /2$.}
\label{qcdsinglepotentialheavy2B}
\end{figure}

\begin{figure}[htb]
\center{\includegraphics[width=6.5cm, angle=90]{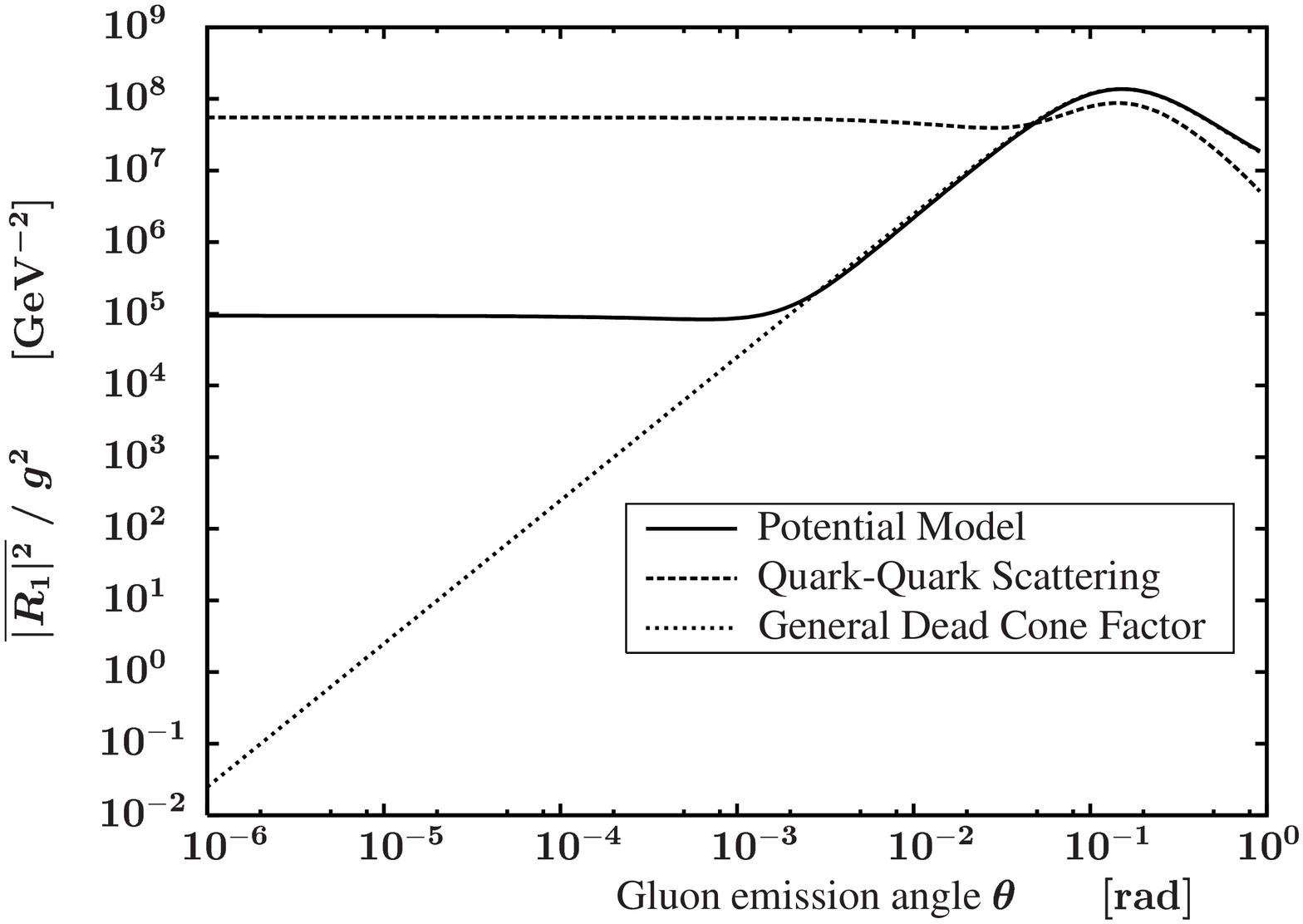}}
\vspace*{-0.5cm} \caption{The radiation amplitude 
$\overline{|R_1|^2} $ as a function of the emission angle $\theta$
is compared for the potential model, the quark-quark scattering on
a light target and the general dead cone factor. The chosen
parameters are $m_d = 0.007$ GeV (light quark), $m_c = 1.5$ GeV (heavy quark),
$\vec{p}_i = (0,0,10)$ GeV, $\vec{p}_{\rm i,Target} = (0,0,0)$, 
$\vec{q}_{\perp} = (0, 0.01)$ GeV, $\mu = 0$, $\omega = 0.001$ GeV,
$\phi = 3\pi / 2$.} 
\label{potentialversusscalarlightprojectile}
\end{figure}

The graphical elements depicted in Fig.~1 correspond to objects in
our C++ code \cite{RT}. Therefore, the matrix elements are
evaluated numerically. All interference effects are exactly dealt
with when squaring sums of matrix elements and performing
averaging/summation over spins/polarisations. Since spinor
calculations do not reveal significant modifications we present
results of the scalar QCD case and comment later on spin effects.

\subsection{Abelian Region}\label{abelianregion}

To highlight the role of the non-abelian diagram 1c with the
triple gluon vertex we consider first the ''abelian part of QCD''.
In Fig.~\ref{combinedheavylightabelian} the radiation amplitude
obtained from the abelian total matrix $M_{\rm 1,rad}^{\rm
abel.QCD}=M_{1,0,0}^{\rm QCD}+M_{1,1,0}^{\rm QCD}$ is shown as a
function of the gluon emission angle $\theta$, which encloses the
directions of the gluon and the incident quark. For the same
kinematics the cases of a light and a heavy projectile are
contrasted and reveal significant suppression of the radiation
pattern at angles $\theta < \theta_0$, which is the dead cone
effect. The heavy particle case is classified with the condition
$|\vec{q}_{\!\perp}| < m$, and vice versa, hence it depends on the
choice of the transverse momentum transfer $\vec{q}_{\!\perp}$.

From analytical and approximated matrix elements
one can derive the $1/(\omega^2\theta^2)$ decrease of the
radiation amplitude for $\theta>\theta_0$ and a
$\theta^2/\omega^2$ slope for $\theta<\theta_0$, as can be seen in
Fig.~\ref{combinedheavylightabelian} for both cases. This
behavior for $\theta \sim \theta_0$ agrees with the dead cone
factor.

Some additional features seen in
Fig.~\ref{combinedheavylightabelian} are clarified in
Fig.~\ref{qcdsinglepotentiallightB}, where the individual
contributions for the light quark situation are depicted,
including the non-abelian contribution. For angles $\theta$
between $\theta_q \equiv \tfrac{|\vec{q}_{\!\perp}|}{E}$ and
$\theta_c \equiv \tfrac{m^2}{\vert \vec{q} \vert E} = \tfrac{\theta_0^2}{\theta_q}$
the pre-emission diagram dominates, and
for $\theta > \theta_q$ pre- and post-emission interfere to yield
a different color factor which accounts for the shift by
$\tfrac{4}{9}$ at $\theta_q$ in
Fig.~\ref{combinedheavylightabelian}. The constant post-emission
process for $\theta < \theta_c$ determines the constant radiation
amplitude in this angular region.

Although in Fig.~\ref{combinedheavylightabelian} 
the picture of the radiation amplitude in the heavy quark
case appears quite similar, here it is the pre- and post emission
which interfere with a magnitude of same order to obtain the dead
cone behavior for $\theta_q < \theta$. The post-emission is
dominant and constant for $\theta < \theta_q$ and thus the
radiation amplitude becomes constant as well.

Note, that in a corresponding QED situation the radiation
probability will be further reduced for $\theta > \theta_q$ due to
destructive interference, which in QCD is excluded by color
factors.

\subsection{Non-Abelian Diagram and Factorization}\label{nonabeliandiagramandfactorization}

We continue to discuss to which degree the three-gluon vertex
diagram 1c contributes and how it may change the behavior of
the abelian QCD single scattering radiation amplitude. In fact,
the results for the kinematical situations considered so far for
abelian diagrams are not effected by the additional three-gluon
vertex diagram. Figure~\ref{combinedheavylightabelian} is not
modified by this contribution; its smallness is demonstrated for
the light quark case in Fig.~\ref{qcdsinglepotentiallightB}.

It can be shown that the non-abelian part becomes especially
important for angles $\theta > \theta_{3glu}$ with
$\theta_{\rm 3glu} \equiv \tfrac{ \vert \vec{q}_{\!\perp} \vert}{\omega}$,
but this implies
$|\vec{q}_{\!\perp}| < \omega,$
in order to consider such an angular configuration.
This is for example realized in Fig.~\ref{qcdsinglepotentialheavy2B} where
calculations of individual contributions in the heavy quark case are shown.
It exhibits that the three-gluon contribution dominates in the whole $\theta$ range.
However,
if one allows for $\theta_{3glu} < 1$, this means to abandon the
soft radiation limit. In other words, the condition $\vert \vec{q}_\perp \vert < \omega$
implies that the momentum transfers, which are constrained by initial and final states,
differ significantly in the radiation and elastic scattering processes. But the latter serves
as reference system for the definition of the radiation amplitude.
The factorization, which is usually  applied in the derivation of
approximated matrix elements, becomes insufficient. In specific examples, this may be
accounted for by an additional correction factor \cite{RT}. Here
we mention that this problem causes e.g. $\theta_0$ in
Fig.~\ref{qcdsinglepotentialheavy2B} to be shifted off the peak
position.

\section{Validity of the Potential Model}\label{validityofthepotentialmodel}

It is tempting to test the validity of the potential model
beyond the assumptions of massless scattering particles.
We analyse this while we contrast the scattering of an incident particle 
in the potential model with
an analogue scattering of the same projectile on a target quark at rest.

It was found that the radiation amplitude calculated in a quark-quark collision
coincides well with the potential model result as long as the angles $\theta$ are small
and the projectile is not heavier than the target.

The results of a scalar QCD calculation for light targets and
heavy projectiles are exhibited in
Fig.~\ref{potentialversusscalarlightprojectile}. There the
radiation amplitude in the potential model can be well-described
by the application of a general dead cone factor, which is valid
for arbitrary angles. For smaller $\theta$ the potential model
result becomes constant (cf.\ Fig.~2) and a simple dead cone
factor does not longer apply. The dead cone suppression is weaker
in a quark-quark scattering approach. This shows that for such an
arrangement of colliding masses the potential model might be
challenged, since the assumption, that no energy would be
transferred to the target, is violated. This is intended to
possibly point to the need of an improved model to describe the
scattering processes in a deconfined medium for
heavy projectile quarks.

Besides this we have numerically justified the usual neglect of radiation contributions
from the target lines in the $A^+$-gauge.
Also comparisons of scalar and spinor calculations
did not reveal significant differences for small angles, light quarks and soft gluons.

In the light quark scattering we find only configurations without
spin flip to contribute to the radiation amplitude.
In contrast, scattering a heavy on a
light quark revealed that configurations become important where
the spin of the light target is flipped.

\section{Conclusions}\label{concl}

This work is aimed  investigating how the mass parameter of an on-shell quark
passing a deconfined medium influences its radiative energy loss probability.
A detailed discussion of the basic single scattering diagrams with
one-gluon emission in the potential model was done.
With the numerical calculation of the radiation amplitude in QCD
we could disentangle interference effects, especially the influence of the
non-abelian effects.

The dead cone suppression factor was shown to emerge in specific angular regions,
however the radiation amplitude deviates from this prediction for either small angles,
due to the post-emission process, or in case of higher gluon energies $\omega$. 
For the latter possibility the three-gluon diagram, 
including the three-gluon vertex, becomes responsible.
We thus have confirmed the suppression effect due to the heavy quark mass, but find the
dead cone factor (2) is not the correct modification
in all kinematical situations.
(This finding agrees with the one in \cite{U.A.W.}, where an averaged dead cone
suppression factor was also found to be insufficient.) 
We stress that not the mass parameter of the projectile itself but
the ratio to the transverse momentum transfer is relevant to categorize light and heavy
mass situations. Non-abelian effects become important
for large gluon energies, that is to say greater than the
transverse momentum transfer.

We have indicated that the potential model might become inappropriate 
if heavy projectile quarks are present. In addition,
it was emphasized that for considerably large values of the gluon energy 
one is confronted with
factorization problems with respect to the elastic scattering part in matrix elements. 
In such situations
the missing unique interpretation of the radiation amplitude questions 
whether it is sufficient to consider
ratios of inelastic and elastic cross sections.

The numerical confirmation of the LPM effect and the calculation
of net energy loss of heavy quarks at finite energy 
along the lines of \cite{U.A.W.,Gyulassy_NPA} are promising
goals of further investigation, such as the extension of our approach to multiple
scattering, i.e.\ higher orders in opacity, and multi-gluon emission.

\section*{Acknowledgement(s)}
Enlightening discussions and guidance of O.P. Pavlenko are
gratefully acknowledged. The work is supported by BMBF 06DR121 and
GSI.

\vfill\eject
\end{document}